\documentclass[journal,twoside,web]{ieeecolor}
\usepackage{IEEE_EDL}
\usepackage{cite}
\usepackage{amsmath,amssymb,amsfonts}
\usepackage{algorithmic}
\usepackage{graphicx}
\usepackage{textcomp}
\def\BibTeX{{\rm B\kern-.05em{\sc i\kern-.025em b}\kern-.08em
    T\kern-.1667em\lower.7ex\hbox{E}\kern-.125emX}}
\graphicspath{{fig/}}

\begin{document}
\title{First demonstration of 200, 100, and 50 $\mu$m pitch Resistive AC-Coupled Silicon Detectors (RSD) with 100\% fill-factor for 4D particle tracking}
\author{M. Mandurrino, R. Arcidiacono, M. Boscardin, N. Cartiglia, G. F. Dalla Betta, \IEEEmembership{Senior Member, IEEE}, M. Ferrero, F. Ficorella, L. Pancheri, \IEEEmembership{Member, IEEE}, G. Paternoster, \IEEEmembership{Member, IEEE}, F. Siviero, M. Tornago
\thanks{Submitted \today; date of current version \today. This work is supported by the Italian National Institute for Nuclear Physics, Gruppo V, through the framework of the RSD Project, by the Horizon 2020 Grants no. UFSD669529 and no. 654168 (AIDA-2020), by the U.S. Department of Energy, Grant no. DE-SC0010107, and by Dipartimenti di Eccellenza, University of Torino (ex L. 232/2016, art. 1, cc. 314, 337). Corresponding author: Marco Mandurrino.}
\thanks{M. Mandurrino and N. Cartiglia are with Istituto Nazionale di Fisica Nucleare, Sezione di Torino, Via P. Giuria 1, 10125 Torino, Italy (email: marco.mandurrino@to.infn.it).}
\thanks{R. Arcidiacono is with CERN, Esplanade des Particules 1, 1211, Meyrin, Switzerland, Universita` degli Studi del Piemonte Orientale, Largo Donegani 2/3, 28100 Novara, Italy, and Istituto Nazionale di Fisica Nucleare, Sezione di Torino, Via P. Giuria 1, 10125 Torino, Italy.}
\thanks{M. Boscardin, F. Ficorella, and G. Paternoster are with Fondazione Bruno Kessler, Via Sommarive 18, 38123 Trento, Italy, and TIFPA-INFN, Via Sommarive 18, 38123 Trento, Italy.}
\thanks{G. F. Dalla Betta and L. Pancheri are with TIFPA-INFN, Via Sommarive 18, 38123 Trento, Italy, and Universita` degli Studi di Trento, Via Sommarive 9, 38123 Trento, Italy.}
\thanks{M. Ferrero, F. Siviero, and M. Tornago are with Istituto Nazionale di Fisica Nucleare, Sezione di Torino, Via P. Giuria 1, 10125 Torino, Italy, and Universita` degli Studi di Torino, Via P. Giuria 1, 10125 Torino, Italy.}}

\maketitle

\begin{abstract}
We designed, produced, and tested RSD (Resistive AC-Coupled Silicon Detectors) devices, an evolution of the standard LGAD (Low-Gain Avalanche Diode) technology where a resistive \mbox{\textit{n}-type} implant and a coupling dielectric layer have been implemented. The first feature works as a resistive sheet, freezing the multiplied charges, while the second one acts as a capacitive coupling for readout pads. We succeeded in the challenging goal of obtaining very fine pitch (50, 100, and 200 $\mu$m) while maintaining the signal waveforms suitable for high timing and \mbox{4D-tracking} performances, as in the standard \mbox{LGAD-based} devices.
\end{abstract}


\section{Introduction}
\IEEEPARstart{R}{SD} (Resistive AC-Coupled Silicon Detectors) are particle detectors with moderate internal gain based on the LGAD (Low-Gain Avalanche Diode) technology~\cite{2014Pellegrini_NIMA,2018Moffat_JINST,2019Giacomini_NIMA}. They are \mbox{\textit{n}-in-\textit{p}} sensors where an additional \mbox{\textit{p}-type} layer, called multiplication or gain layer, is implanted under the \mbox{\textit{n}-type} cathode. In reverse bias conditions, such a layer provides an electric field which is responsible for the avalanche impact ionization of charge, as occurring in standard APD~\cite{2007Sze}, where the multiplication factor is several orders of magnitude higher. Since the readout amplification would enhance also the electronic noise, having an internal gain is beneficial for timing applications because it mainly provides signal multiplication. Indeed, when the gain is not too high, the S/N (signal to noise) ratio increases and, if the gain is high enough, the slew rate increases as well. Both the S/N and the slew rate are strictly correlated to the time resolution $\sigma_\text{t}$~\cite{2017Cartiglia_NIMA}, and several simulations and measurements demonstrated that, for high-energy charged particles detection, a moderate gain (\mbox{$G \sim$~10--20}) in thin sensors (\mbox{$\sim 50 \, \mu$m}) allows to reach \mbox{$\sigma_\text{t}\sim$~30~ps}~\cite{2015Cartiglia_NIMA,2018Sadrozinski_RPP}. Another key aspect of using the LGAD technology is the \mbox{well-known} behavior of its radiation tolerance: as it has been already proven on UFSD (\mbox{Ultra-Fast} Silicon Detectors, i.e. LGAD optimized for timing) this technology can operate in high-radiation environments, with fluence $\phi$ as large as \mbox{$\phi \sim 10^{15}$~n$_{\text{eq}}$/cm$^2$}~\cite{2019Ferrero_NIMA}, which is fully compatible with the value expected in the MIP Endcap Timing Layer (ETL) for the \mbox{High-Luminosity} upgrade of CMS at CERN.

\begin{figure}[!h]
\centerline{\includegraphics[width=\columnwidth]{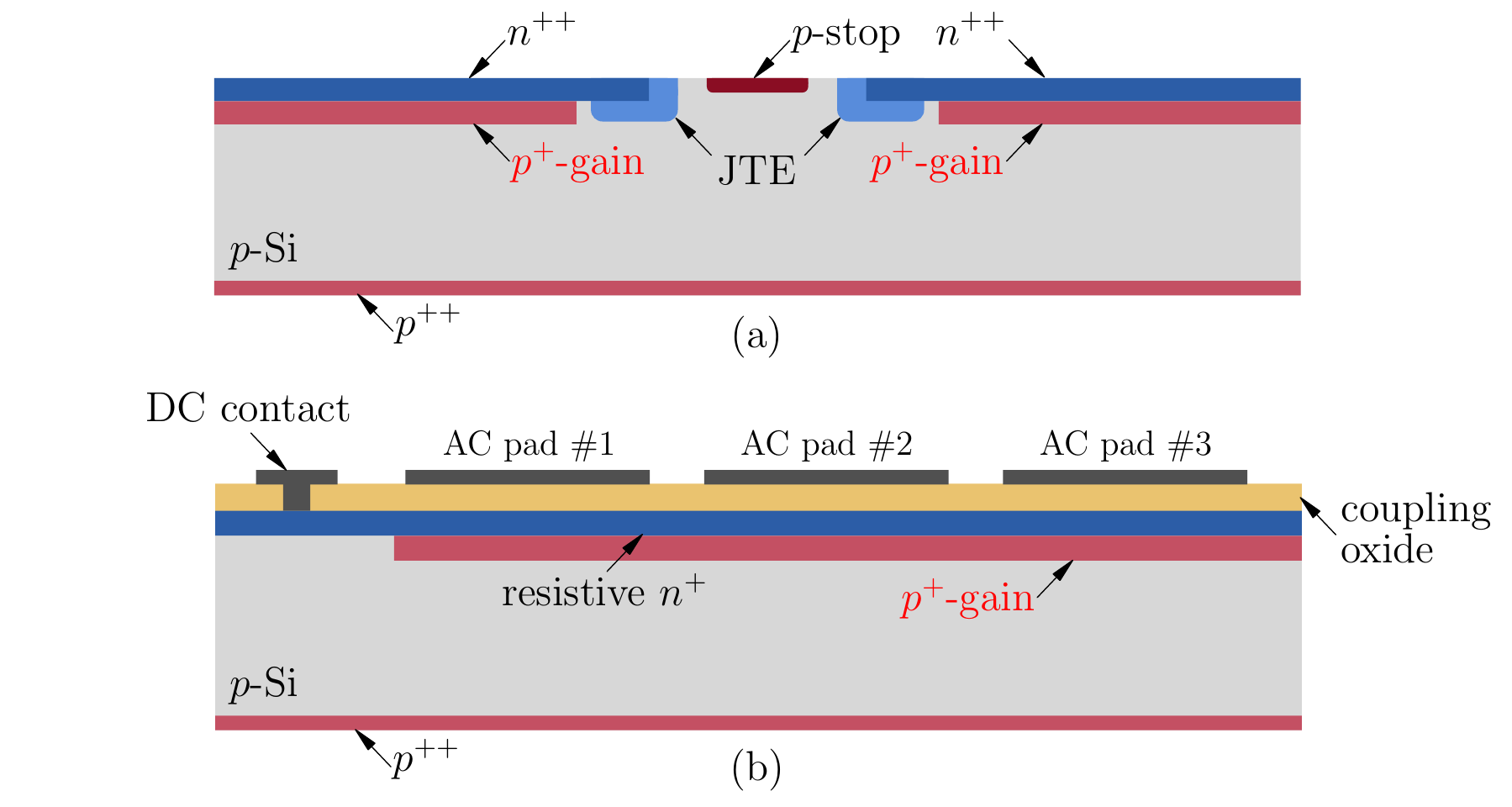}}
\caption{Cross-section of an LGAD (a) in correspondence of the \mbox{inter-pad} region -- where the JTE (Junction Termination Extension) and the \mbox{\textit{p}-stop} are the segmentation implants -- and of an RSD (b).}
\label{fig:crosssections}
\end{figure}

In pixelated LGAD, the electrode segmentation requires to include additional implants, namely the JTE (Junction Termination Extension), aimed at avoiding early breakdown at the edge, and \mbox{\textit{p}-stop}, aimed at electrically isolating neighbour pixels at the surface, as reported in subfigure~\ref{fig:crosssections}~(a).
This strategy decreases the geometrical efficiency of the detector, because the segmentation structures introduce a discontinuity in the multiplication mechanism and, thus, a loss of efficiency in the track reconstruction. This loss is strictly related to a decreased \mbox{fill-factor} (the ratio between the active and the total area) that, in \mbox{state-of-the-art} \mbox{LGAD-based} trackers, typically ranges from 70 to 90\%, depending on the sensor pitch and on the foundry's technology~\cite{2017Mandurrino_31stRD50}.

The next generation of particle detectors requires the capability of performing 4D tracking~\cite{2015Cartiglia_NIMA} with a very high time resolution, spatial granularity and almost perfect geometrical acceptance. This holds either to assign correctly particles belonging to the same interaction or to perform particle identification using \mbox{time-of-flight} technique. As an example, due to the hermetic coverage needed in the ETL, the \mbox{fill-factor} of its trackers must be $>95 \%$~\cite{ETLproposal,2018Bendavid}. To achieve such result we designed, produced, and tested a run of RSD particle sensors~\cite{2018Mandurrino_NSS,2019Mandurrino_34thRD50} within a collaboration among the Torino division of INFN (the Italian National Institute for Nuclear Physics), the University of Trento, and Fondazione Bruno Kessler (FBK).

RSD are \mbox{LGAD-based} devices with unsegmented gain layer, which spreads throughout all the sensor area, which means a 100\% \mbox{fill-factor}. The readout segmentation is obtained at the level of the AC metal pads, which are capacitively coupled to the detector bulk via a dielectric spacer layer deposited between the Silicon and the readout. The second working principle at the basis of RSD is the implementation of a \mbox{\textit{n}-type} resistive implant beneath the coupling layer which allows (\emph{i}) the induction of the signal in the AC metal pads and (\emph{ii}) to control the slow discharge of the multiplied charges with a time constant given by the RC of the equivalent circuit. In particular, the discharge process is required to be sufficiently long to guarantee the complete signal induction, but also short enough to minimize pile-up effects~\cite{2018Mandurrino_NSS}. Therefore, the resistive layer and the capacitive oxide act as an RC circuit in the readout path.

When the multiplied charges drift towards the electrodes, they induce a first (positive) signal on the pads, whose amplitude is proportional to C. Then, the electrons reach and flow through the resistive sheet, where they are slowed down during their discharge to the DC contact. This induces a second (negative) signal, or undershoot, whose duration is proportional to R. By maintaining the duration of the first lobe as in the signals coming from standard LGAD-based detectors, and designing the RC in such a way that both the amplitude and the discharge time of the undershoot are properly optimized, RSD allow to minimize pile-up effects without losing in rate capabilities.

\section{Devices production and testing}
The spatial resolution in RSD is essentially determined by the pitch of the readout AC pads, as it occurs for the channels of a generic Silicon detector. For this reason we included in the first production run, named RSD1, various detector geometries, pitch and pad sizes, from more relaxed \mbox{500~$\mu$m~$\times$~500~$\mu$m} pitch square matrices to the finest ones with \mbox{50~$\mu$m~$\times$~50~$\mu$m} pitch. The goal is to find the best \mbox{AC-related} parameters, i.e. the dielectric thickness and the \mbox{\textit{n}-layer} resistivity, which allow a proper reconstruction of the particle hit position in each of these designs.

To explore the phase-space of the RSD parameters, we processed 15 6$^{\prime\prime}$ Silicon wafers, both Float Zone (FZ) and Epitaxial (Epi), with several splits of the most important implants. In particular, the Boron dose in the gain layer spreads from 0.92 to 0.96, in steps of 0.02 each, where such numbers are in the same normalized scale of those reported in Table~1 of~\cite{2019Ferrero_NIMA}. For what concerns the \mbox{\textit{n}-type} cathode, we used a standard implantation dose (named B) which is approximately a tenth of that one used in all LGAD presented in~\cite{2019Ferrero_NIMA}, plus two other splits, where this dose is halved (A) and doubled (C), respectively. Moreover, the dielectric layer has been implemented in two configurations, low (L) and high (H), by slightly changing the growth thickness during the deposition process to test different values of coupling capacitance on the AC pads. Finally, two types of wafers (FZ and Epi) have been processed in order to understand if the intrinsic difference in their bulk resistivity (respectively \mbox{$>$\,3\,k$\Omega$\,cm} and \mbox{$>$\,1\,k$\Omega$\,cm}) may play a role in the devices operation, and to test the compatibility of  different substrate technologies with the RSD fabrication process.

After characterizing the production run at FBK through $I(V)$ and $C(V)$ measurements, which have shown a high homogeneity in terms of implantation doses, both within each wafer and among different wafers, the devices have been diced and then characterized in Torino with a IR laser (\mbox{$\lambda \sim 1064$~nm}) in a \mbox{front-TCT} (Transient Current Technique) equipment~\cite{1993Eremin_NIMA}. The laser illumination comes from the front side of the detector and, through ionization, generates a certain number of primary charges that undergo multiplication mechanism whose magnitude depends on the laser optical power, on the applied bias, and on the Boron dose in the gain layer.

All RSD devices have a DC collecting electrode surrounding the matrix of AC pads. Such \mbox{DC-contact} is grounded, while the reverse polarization is imposed by the voltage on the back side. The detector guardring (GR), collecting the leakage current coming from the device edges, is also connected to ground. Our testing campaigns involved several structures. Here we focus on the most challenging designs: the matrices with \mbox{3$\times$3} pads of \mbox{$200~\mu$m$~\times~200~\mu$m}, \mbox{$100~\mu$m$~\times~100~\mu$m}, and \mbox{$50~\mu$m$~\times~50~\mu$m} pitch. The structures tested have \mbox{$150~\mu$m$~\times~150~\mu$m}, \mbox{$70~\mu$m$~\times~70~\mu$m}, and \mbox{$35~\mu$m$~\times~35~\mu$m} AC pad size, respectively. All the contacts have been realized through \mbox{wire-wedge-bonding} and the signals were read by broadband amplifiers (40~dB from CIVIDEC~\cite{cividec}) concurrently from three AC pads while the others have been grounded, when possible.

\section{Results}
\begin{figure}[!h]
\centerline{\includegraphics[width=\columnwidth]{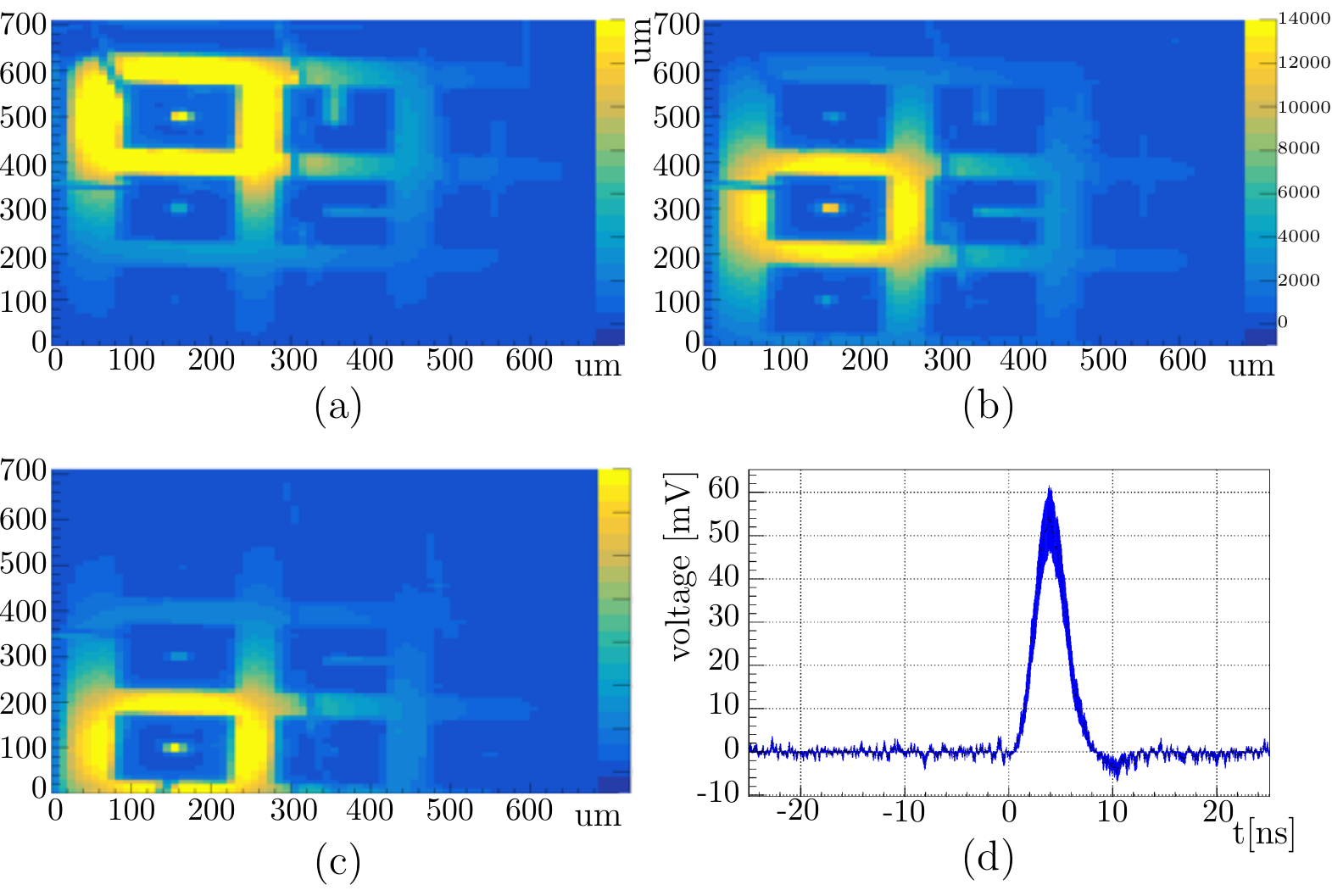}}
\caption{2D maps of induced charge obtained with a TCT setup after an integration time of about 5~ns for three different AC pads (a)--(c) of a 200~$\mu$m pitch RSD from W10 at 250~V (a typical value for a thin \mbox{LGAD-based} detector to operate); from the maps it is evident the presence of some optical windows (slits/spots) on the metal readout pads. Signal waveform produced by the same device (d).}
\label{fig:map_200}
\end{figure}

The first device we present is the \mbox{3$\times$3} pads matrix with \mbox{$200~\mu$m$~\times~200~\mu$m} pitch and \mbox{$150~\mu$m$~\times~150~\mu$m} pad size coming from wafer~10 (substrate: FZ, \textit{n}$^+$ dose: B, \textit{p}$^+$ dose: 0.96, dielectric thickness: H). After calibrating the laser focus, reaching a final spot with a diameter of about 15~$\mu$m, the TCT laser scanned all the active part of the sensor, which has an area of \mbox{$\sim 0.44$~mm$^2$}. By integrating the AC signals coming from the three active channels for an integration time of about 5~ns, we obtained the 2D maps of charge reported in Figure~\ref{fig:map_200}. It is worth noting that, for this sensor (that, at 250~V, has a gain $\sim$~15--20) as well as for the other structures here presented, we set a laser intensity which produces $\sim 1$~MIP.

As one may see, each channel shows a collection volume around its corresponding pad. This means that the overlap of neighboring collecting zones is such that it is possible to reconstruct the particle/laser hit position by weighting the number of charges induced on each channel of that cluster. Moreover, all the maps show the presence of the optical windows inside the AC pads, either square openings (as in the leftmost column) or slits with different orientations, that we designed in order to perform timing characterizations or specific \mbox{inter-pad} laser scans.

\begin{figure}[!h]
\centerline{\includegraphics[width=\columnwidth]{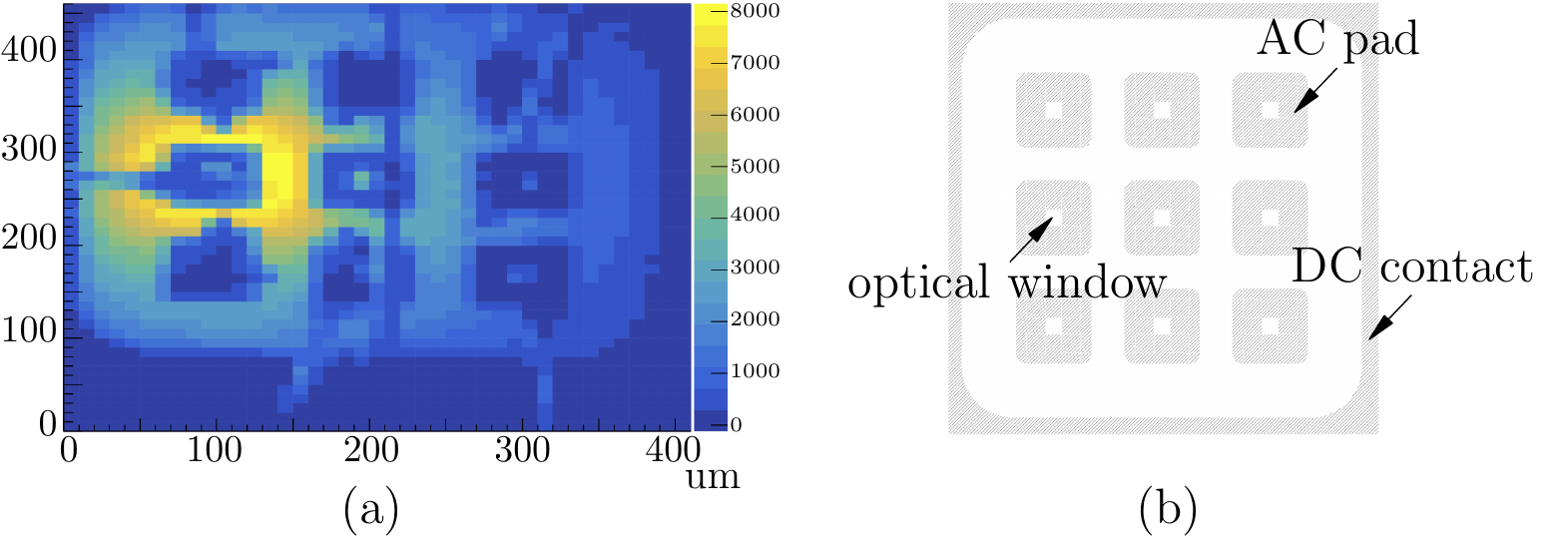}}
\caption{2D maps of induced charge in a 100~$\mu$m pitch RSD from W10 (a) obtained with a TCT setup at 250~V for a time integration of 3~ns and layout of the device (b).}
\label{fig:map_100}
\end{figure}

\begin{figure}[!h]
\centerline{\includegraphics[width=\columnwidth]{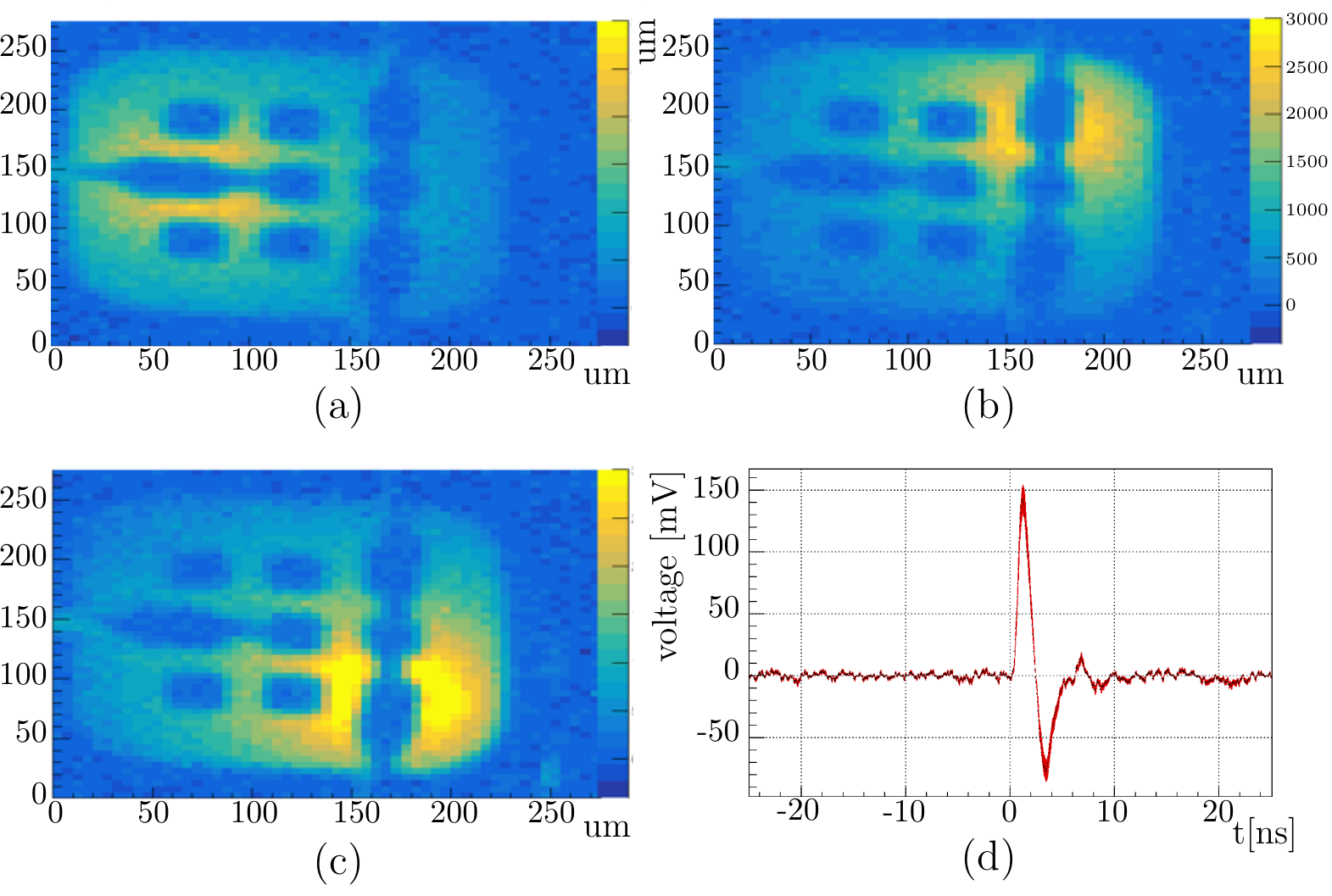}}
\caption{2D maps of induced charge in a 50~$\mu$m pitch RSD from W8 (a)--(c) obtained with a TCT setup at 300~V and by integrating the AC signals for 2.7~ns. Signal waveform produced by the same device (d).}
\label{fig:map_50}
\end{figure}

To study the 100 and 50~$\mu$m pitch RSD with the same laser setup, we measured samples from, respectively, wafers~10 and 8 (substrate: FZ, \textit{n}$^+$ dose: B, \textit{p}$^+$ dose: 0.94, dielectric thickness: L). The resulting maps of charge, where the integration time is 3~ns for the 100~$\mu$m and 2.7~ns for the 50~$\mu$m pitch, are reported in Figures~\ref{fig:map_100} and \ref{fig:map_50}. In order to investigate the segmentation properties of RSD1, for the 100~$\mu$m pitch sample we also took data with two floating (without \mbox{wire-bond}) AC pads and the signal did not degrade. For the 50~$\mu$m pitch detector, we also tested a sensor with three AC pads connected and six pads floating. As in the previous case, no apparent differences have been observed with respect to the sensor with three pads connected and six pads grounded (through \mbox{wire-bonding}).

The effects of different RC on the signal shape can be seen comparing Figures~\ref{fig:map_200}(d) and \ref{fig:map_50}(d): consistently with the RSD working principle, already explained in the introduction, and with our simulations~\cite{2018Mandurrino_NSS,2019Mandurrino_34thRD50}, for higher RC the first (positive) lobe dominates on the undershoot component. This is because the R are the same in both wafers (i.e., same \textit{n}$^+$ dose) but the C are different (from simulations, the coupling capacitance is $\sim 3.5$~pF in the 200~$\mu$m pitch detector and $\sim 0.3$~pF in the 50~$\mu$m). The difference in peak intensities is essentially due to the different bias voltage at which the two RSD are operating (250 and 300~V, respectively). Nevertheless, even not at the same biasing conditions, the integral of the first lobe is higher in the RSD with the highest coupling capacitance.

\section{Conclusion}
To circumvent the issues in realizing a particle detector for \mbox{4D-tracking} with high spatial/timing resolution, the highest acceptance area as possible, and with expected \mbox{radiation-hardness} as in LGAD, we designed, produced, and tested a first batch of devices based on the RSD paradigm, an evolution of the traditional LGAD technology. In addition to few other groups, that have recently reported the fabrication of similar detectors~\cite{2017Sadrozinski_RD50,2019Giacomini_arXiv}, in this work we reported the first evidence of a working \mbox{ultra-fine-pitch} segmentation in AC-coupled Silicon particle detectors with internal gain and 100\% \mbox{fill-factor}. Moreover, by characterizing devices with 200, 100, and 50 $\mu$m pitch through the TCT pulsed IR laser, we demonstrated that it is possible to control the shape of the (bipolar) waveform by choosing the parameters of the RSD design and process. In particular, high RC in the readout path correspond to a first (positive) lobe dominating on the second (negative) undershoot, making this kind of waveforms very similar to the unipolar signals of standard \mbox{LGAD-based} detectors, that are already optimized for \mbox{4D-tracking}.

\end{document}